\documentclass[runningheads]{llncs}
\usepackage{graphicx}
\usepackage{listings}
\usepackage{svg}
\usepackage{hyperref}
\usepackage{multirow}
\usepackage{tabularx}
\usepackage{longtable}
\usepackage{adjustbox}
\usepackage{enumitem}
\usepackage{amsfonts}
\begin{document}
\definecolor{lightBlue}{RGB}{0, 153, 255}
\definecolor{lightRed}{RGB}{204, 0, 255}
\def\speciallstcolor{\begingroup\color{green}}
\def\endspeciallstcolor{\endgroup}
\lstdefinelanguage{Manchester}
{
    sensitive = true,
    keywords = [1]{Class, EquivalentTo, SubClassOf},
    morekeywords = [4]{exam, value, patient, profile, sys, sysValue, dias, diasValue, plan, ethnicity, demo, NewTreatmentSubPlan, RecommendDiabetesScreening, h, w, h_exp, bmi},
    keywordstyle=[3]\color{lightRed}\textbf,
    keywordstyle=[4]\color{lightBlue}\textbf,
    morestring=[b]'',
    alsoletter=:
}
\title{Process Trace Querying using Knowledge Graphs and Notation3}
%
%
\author{William Van Woensel\orcidID{0000-0002-9421-8566}}
%
\authorrunning{W. Van Woensel}
%
\institute{University of Ottawa, Ottawa, ON K1N 6N5, Canada
\\
\email{wvanwoen@uottawa.ca}}
\maketitle              
\begin{abstract}
In process mining, a \textit{log exploration step} allows making sense of the event traces; e.g., identifying event patterns and illogical traces, and gaining insight into their variability. 
To support expressive log exploration, the event log can be converted into a Knowledge Graph (KG), which can then be queried using general-purpose languages.
We explore the creation of semantic KG using the Resource Description Framework (RDF) as a data model, combined with the general-purpose Notation3 (N3) rule language for querying.
We show how typical trace querying constraints, inspired by the state of the art, can be implemented in N3.
We convert case- and object-centric event logs into a trace-based semantic KG; 
OCEL2 logs are hereby ``flattened'' into traces based on object paths through the KG.
This solution offers (a) expressivity, as queries can instantiate constraints in multiple ways and arbitrarily constrain attributes and relations (e.g., actors, resources); 
(b) flexibility, as OCEL2 event logs can be serialized as traces in arbitrary ways based on the KG; 
and (c) extensibility, as others can extend our library by leveraging the same implementation patterns.

\end{abstract}
\section{Introduction}

In practice, process mining typically includes an intermediate \textit{exploration step} between event log pre-processing and the actual process discovery, which aims to properly understand the event traces. 
For instance, in a healthcare-related methodology for process diagnostics~\cite{REBUGE201299}, it is part of \textit{log inspection}; in the PM\textsuperscript{2} methodology, it is part of \textit{data processing}~\cite{Eck2015PM}. 
This step may point out unexpected event patterns, which allow for more pointed questions to the process owner (e.g., on activity meanings); identify illogical traces that occur due to poor data quality; 
or help to gain insight into the variability of the event log, e.g., for separating them into distinct subsets for mining. 
Importantly, by better understanding the event log, one may avoid so-called spaghetti models~\cite{pm_handbook} that often result from the blind application of process discovery on event logs. 

Process mining tools can be used for this step, which typically offer some graphical log exploration support (such as a variant explorer or variant querying feature~\cite{Schuster-process_querying}). 
Alternatively, tailored domain-specific languages (DSL) have been developed for more expressive querying of process variants or event traces~\cite{Schuster-process_querying,Beheshti-process_querying}.
We explore the creation of a semantic Knowledge Graph (KG) from an event log, called an Event Log KG (ELKG)\footnote{This term was chosen to differentiate from Event Knowledge Graphs (EKG)~\cite{Fahland2022}.} which can then be queried for log exploration.
To support trace querying constraints, we target a trace-based KG, i.e., where the event log is stratified into a series of event traces organized on case notions.
We implemented a set of trace constraints on the (non)-occurrence of activities and sequential relations~\cite{Schuster-process_querying,Beheshti-process_querying}. 
We further show how case- and object-centric event logs can be converted into an ELKG. 
OCEL2 logs, which lack a central case notion and can instead associate events with multiple objects, are ``flattened'' into traces by leveraging the events' associated object graph in the ELKG.

We use the graph-based Resource Description Framework (RDF)~\cite{cyganiak_rdf} as the KG data model, and Notation3 (N3)~\cite{van_woensel_notation3_2023} as a general-purpose rule language for reasoning over and querying the KG.
N3 allows navigating graph-based data and offers a variety of built-ins for logical, string and numeric operations and comparisons.
We implemented a set of state-of-the-art trace constraints~\cite{Schuster-process_querying,Beheshti-process_querying} via recursive backward-chaining rules in N3. 
Using these constraints, one can thus use N3 to perform trace querying during process exploration.
As an extra benefit, our work serves to illustrate how to implement trace querying in a rule language. 
Our contribution offers \textit{expressivity}, as trace constraints can be instantiated in multiple ways, and any arbitrary constraint can be issued on activities, events, and traces; \textit{flexibility}, as OCEL2 logs can be converted into sets of traces in different ways, by leveraging the ELKG; and \textit{extensibility}, as others can implement additional constraints to suit their goals based on our implementation patterns.
Our implementation is available online~\cite{elkg_repo}.

This paper is structured as follows. We start by discussing preliminaries (Section~\ref{sec:prelim}). Subsequently, we elaborate on converting event logs into ELKG (Section~\ref{sec:conv_logs_elkg}) and implementing trace querying constraints (Section~\ref{sec:query_process_traces}). We present a preliminary performance evaluation (Section~\ref{sec:eval}) and compare with the state of the art (Section~\ref{sec:rel_work}). We finish with conclusions and future work (Section~\ref{sec:concl_fut_work}).


\section{Preliminaries}
\label{sec:prelim}

\subsection{Case- and Object-Centric Event Logs}
\label{sec:ccel_ocel}
\textbf{Case-Centric Event Logs (CCEL)}.
A ``traditional'' CCEL~\cite{pm_handbook}, which is typically serialized using the Extensible Event Stream (XES) format~\cite{xes_2023}, is defined as a tuple $L = (E, \#, \prec)$, where the set of events $E \subseteq \mathbb{U}_{ev}$ (universe of events); mapping $\# \in E \rightarrow \mathbb{U}_{map}$ (universe of functions $U_{att} \rightarrow U_{val}$), which is used to assign values to event attributes; and a total ordering~\footnote{For the purposes of this paper, we consider a \textit{total} ordering between events.}
$\prec \subseteq E \times E$ of events based on their timestamps, where $e_k \prec e_l$ iff $\#(e_k)(time) < \#(e_l)(time)$.

A trace groups all events that share the same chosen ``case'' (e.g., purchase, visit), sorted ascending on their timestamp. 
We define a trace~\cite{pm_handbook} as $\sigma = <e_1, e_2, ..., e_n> \in E^*$, where events $e_i$ share the same case attribute $\#(e)(case)$ and are sorted on their total ordering
$\prec$.
Based on the concept of a trace and the $\prec$ relation, we define a \textit{sequential relation} $\rightarrow \subseteq E \times E$ as follows:
$e_k \rightarrow e_l$ iff $\exists \sigma \in L: e_k, e_l \in \sigma \wedge e_k \prec e_l \wedge \neg \exists e_i \in \sigma: e_k \prec e_i \wedge e_i \prec e_l$.
I.e., $e_k$ and $e_l$ are part of the same trace $\sigma$ where $e_k$ occurs before $e_l$, with no other event in between.

\vspace{0.2cm}
\noindent \textbf{Object-Centric Event Log (OCEL).}
An OCEL~\cite{berti_ocel_2023} is defined as a tuple $L = (E, O, EA, OA, evtype, time, objtype,
eatype, oatype, eaval, oaval, E2O,O2O)$.\\
For our purposes, we focus on the following elements:
$E \subseteq \mathbb{U}_{ev}$ is the set of events; 
$O \subseteq \mathbb{U}_{obj}$ is the set of objects;
$evtype: E \rightarrow \mathbb{U}_{etype}$ assigns types (activities) to events; 
$objtype: O \rightarrow \mathbb{U}_{otype}$ assigns types to objects;
$time: E \rightarrow \mathbb{U}_{time}$ assigns timestamps to events;
$E2O \subseteq E \times \mathbb{U}_{qual} \times O$ are qualified event-to-object relations;
and $O2O \subseteq O \times \mathbb{U}_{qual} \times O$ are qualified object-to-object relations.

Note that, instead of organizing events around a single case, events can be associated with multiple objects ($E2O$).
This avoids the typical convergence and divergence problems of CCEL~\cite{berti_oc-pm_2023}.
Nevertheless, to analyze an OCEL using current techniques (e.g., process discovery), it is typically ``flattened'', i.e., converted into a traditional CCEL, by choosing a related object as a case notion~\cite{berti_oc-pm_2023}. 

\subsection{Resource Description Framework (RDF) and Notation3 (N3)}
\label{sec:rdf_n3}
\textbf{RDF}~\cite{cyganiak_rdf},
the building block of the Semantic Web, 
describes information as \textit{triples} with subject, predicate and object terms. This gives rise to a graph structure, with subjects and objects as nodes and predicates as edges.
Terms include URIs, which identify resources (entities); blank nodes, for resources lacking an identifier;
and literals (e.g., numbers, strings). 
An example RDF snippet:

\lstset{language=Manchester, basicstyle=\ttfamily\fontsize{9}{10}\selectfont, 
columns=fullflexible, xleftmargin=0mm, framexleftmargin=5mm, breaklines=true, breakatwhitespace=false, numberstyle=\ttfamily\fontsize{9}{10}\selectfont, numbersep=5pt, tabsize=2, mathescape=true, literate= {==>}{$\rightarrow{}$}{1},showstringspaces=false,escapeinside={(*@}{@*)}}

\begin{lstlisting} 
p2p:event1 tr:activity p2p:create_purchase_requisition .
p2p:event1 tr:timestamp "2014-10-22T09:27:00+00:00" ;
    p2p:initiator _:b1 .
_:b1 p2p:employeeName "employee A" .
\end{lstlisting}

URIs are shown as qualified names (e.g., \textit{tr:activity}) that include a namespace (\textit{tr}) and localname (\textit{activity}). Namespace definitions for \textit{tr} and \textit{p2p} are not shown for brevity.
Semicolons ``;'' can be used to group triples with the same subject (e.g., \textit{p2p:event1}). Blank nodes are shown as \verb|_:b<nr>|.

\vspace{0.2cm}
\noindent \textbf{N3}~\cite{van_woensel_notation3_2023} is a rule language for reasoning over and querying RDF. 
N3 rules are expressed using triples with variables (e.g., ``?x''), lists written using ``()'',
and graph terms using ``\{\}'' to group triples as rule heads and bodies.
E.g., this rule will infer that it is cloudy, since we assert that it is raining:

\lstset{language=Manchester, basicstyle=\ttfamily\fontsize{9}{10}\selectfont, 
columns=fullflexible, xleftmargin=0mm, framexleftmargin=5mm, breaklines=true, breakatwhitespace=false, numberstyle=\ttfamily\fontsize{9}{10}\selectfont, numbersep=5pt, tabsize=2, mathescape=true, literate= {==>}{$\rightarrow{}$}{1},showstringspaces=false,escapeinside={(*@}{@*)}}

\begin{lstlisting}
wr:weather rdf:type wr:Raining .
{ wr:weather rdf:type wr:Raining } => { wr:weather rdf:type wr:Cloudy } .
\end{lstlisting}

Rules can be either forward-chaining (using \verb|=>|), meaning that inferences are directly materialized in the KG; or backward-chaining (using \verb|<=|), meaning that they can help to resolve queries or other rules (cfr. logic programming). 

\section{Converting Event Logs into ELKG}
\label{sec:conv_logs_elkg}

\subsection{Case-Centric Event Logs (CCEL)}
\label{sec:conv_ccel}

We represent a CCEL as an RDF-based ELKG with the following sets of triples.
We write $\langle s, p, o \rangle$ to represent an RDF triple.
For a log entity such as a trace $\sigma$, we write $\sigma^{t}$ for the corresponding RDF term. 
This term may be a URI, blank node, or literal, depending on the entity.

\vspace{0.2cm}
\noindent (1) $\{\, \langle \sigma^{t}, type, Trace\rangle , \langle e^t, in, \sigma^t \rangle \mid \sigma \in L, e \in \sigma \}$\\
(2) $\{\, \langle e^{t}, att^{t}, val^{t}\rangle  \mid e \in E \wedge att \in dom(\#(e)) \wedge val = \#(e)(att) \}$\\
(3) $\{\, \langle e_{k}^t, next, e_l^t \mid \sigma \in L \wedge e_k, e_l \in \sigma \wedge e_k \rightarrow e_l \}$\\
(4) $\{ \, \langle e^t, next, \emptyset \rangle, \mid \sigma \in L \wedge e \in \sigma \wedge \neg \exists e_n \in \sigma: e \prec e_n \}$

\vspace{0.2cm}
\indent For each trace $\sigma$ in log $L$, triples describe $\sigma^{t}$ with type \textit{Trace}, and all its events $e^t$ as being part of (\textit{in}) $\sigma^t$ (1).
For each event and their attribute \textit{att}, 
a triple describes the event $e^{t}$ using $att^{t}$ and its $value^{t}$ (2). 
Each sequential relation $e_k \rightarrow e_l$ is described by a triple linking the prior event $e^t_k$ with the following event $e^t_l$ using predicate $next$ (3).
We further keep an explicit end event $nil$ (4).

Below, we show a simplified snippet of the Sepsis event log~\cite{mannhardt_sepsis} in RDF:

\lstset{language=Manchester, basicstyle=\ttfamily\fontsize{9}{10}\selectfont, 
columns=fullflexible, xleftmargin=5mm, framexleftmargin=5mm, numbers=left, stepnumber=1, 
breaklines=true, breakatwhitespace=false, numberstyle=\ttfamily\fontsize{9}{10}\selectfont, numbersep=5pt, tabsize=2, frame=lines, captionpos=t, mathescape=true, literate= {==>}{$\rightarrow{}$}{1},showstringspaces=false,escapeinside={(*@}{@*)},numberbychapter=false, 
caption={RDF snippet of the Sepsis event log}, label=lst:log_rdf}

\begin{lstlisting} 
@prefix tr: <http://notation3.org/trace#> .
@prefix se: <http://dutch.hospital.nl/sepsis#> .

se:event1 tr:activity se:ER_Registration ; tr:in se:trace_A ;
    tr:timestamp "2014-10-22T09:15:41+00:00" ;
    se:InfectionSuspected true . # ...

se:event1 tr:in se:trace_A . se:event2 tr:in se:trace_A .
se:event1 tr:next se:event2 . se:event2 tr:next rdf:nil .
\end{lstlisting}

\textit{Event1} has activity attribute \textit{ER Registration}, a timestamp and other attributes (e.g., \textit{InfectionSuspected}).
Sequential event relations are described between \textit{event1} and \textit{event2}, and with the explicit end event \textit{nil}.

\subsection{Object-Centric Event Logs (OCEL2)}
\label{sec:conv_ocel}

We represent an OCEL2 as an RDF-based ELKG with the following triples (focusing on the elements from Section~\ref{sec:ccel_ocel}), using the same notation as before:

\vspace{0.2cm}
\noindent (1) $\{ \langle e^t, activity, a^t\rangle, \langle e^t, timestamp, t^t \rangle \mid \forall e \in E: a=evtype(e), t=time(e) \}$\\
(2) $\{ \langle o^t, type, t \rangle \mid \forall o \in O: t=objtype(o) \}$\\
(3) $\{ \langle e2o^t, event, e^t\rangle, \langle e2o^t, object, o^t\rangle, \langle e2o^t, qualifier, q^t\rangle \mid (e, q, o) \in E2O \}$\\
(4) $\{ \langle o2o^t, object, o^t\rangle, \langle o2o^t, object2, o_2^t\rangle, \langle o2o^t, qualifier, q^t\rangle \mid (o, q, o_2) \in O2O \}$

\vspace{0.2cm}
Per event \textit{e}, we keep its activity and timestamp (also lifecycle and resource; not shown) (1); for each object, we keep its type (2). 
We \textit{reify} \textit{E2O} and \textit{O2O} relations and use triples to describe their events, objects and qualifiers (3)-(4).
To analyze an OCEL using current techniques (e.g., process discovery), the log is typically ``flattened''~\cite{berti_oc-pm_2023}, i.e., converted into a CCEL.
This is done by choosing a particular object type $ot$ as a case notion.
An individual object \textit{o} with type \textit{ot} thus becomes a case, in CCEL terms, with its associated trace including all of its related events, i.e., $\{ e \in E \mid objtype(o)=ot \land (e, q, o) \in E2O \}$.


This flattening operation does not cover larger case perspectives. 
Consider an OCEL2 log on a P2P (procure-to-pay) process, which we want to flatten into traces that cover all Purchase Requisition (PR) events from a buyer's perspective;
i.e., from the PR's creation, quotations, related purchase orders and receipts, to payment.
Such a perspective allows e.g., checking for long periods from PR creation to ultimate payment; duplicate payments for a given PR; or analyze events that follow ``maverick buying'' (i.e., PR's without approval).

Choosing the ``PR'' object type as the case notion would not cover all these events.
Fig~\ref{fig:p2p-graph} shows a simulated P2P case by Park et al.~\cite{p2p_case}, 
where the PR object type is only linked to a subset of relevant activities:

\begin{figure}
    \centering
    \includegraphics[width=0.8\linewidth]{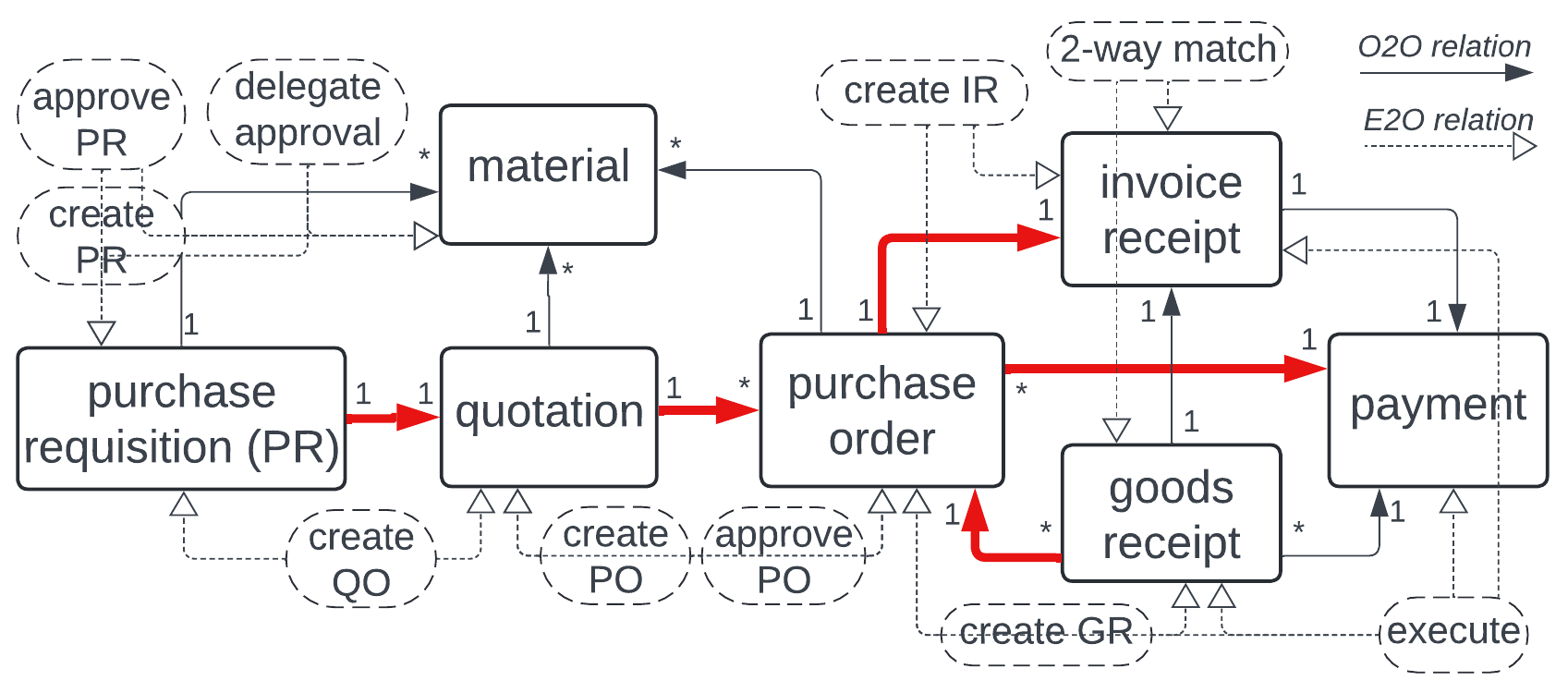}
    \caption{Object types and event activities for a P2P use case. Object types are rectangles with O2O links between them; event activities are dashed ellipses with E2O links to object types. Red arrows represent an object path that collects all relevant events.}
    \label{fig:p2p-graph}
\end{figure}

Indeed, E2O relationships state that an object affects an event or vice-versa~\cite{berti_ocel_2023}; they will likely not link an object  (e.g., PR) to all events relevant for our purposes.
To fill in these gaps, we can leverage O2O relations: starting from an individual PR object, the highlighted O2O path identifies objects with E2O relations to all relevant events, i.e., from PR creation to payment~\footnote{Here, the path captures most object types and activities in the log; this will not always be the case, depending on the log's scale and desired perspective.}.

We define a $\textrm{\textit{Perspective}}(ot_{st}, Path_{OG})$
with start object type $ot_{st}$ and path $Path_{OG}$ through the \textit{O2O} object graph, called $OG$.

A path $Path_{OG}$ is a tuple $(Step_{OG}^0, ..., Step_{OG}^n)$ 
where a step $Step_{OG} \in \mathbb{U}_{Step_{OG}} = \mathbb{U}_{Rel_{OG}} \; \cup \; \mathbb{U}_{Alt_{OG}} \; \cup \; \mathbb{U}_{Path_{OG}}$:\\
\indent $\mathbb{U}_{Rel_{OG}}: \{ \; \leftarrow,\rightarrow \; \} \; \times \; \mathbb{U}_{qualifier}$ where $Rel_{OG}$ is a tuple \textit{(direction, qualifier)};\\
\indent $\mathbb{U}_{Alt_{OG}}: \mathcal{P}(\mathbb{U}_{Step_{OG}})$ where $Alt_{OG}$ is a tuple $(Step_{OG}^0, ..., Step_{OG}^n)$;\\
\indent $\mathbb{U}_{Path_{OG}}: \mathcal{P}(\mathbb{U}_{Step_{OG}})$ where $Path_{OG}$ is a tuple $(Step_{OG}^0, ..., Step_{OG}^n)$.

\vspace{0.3cm}
The $collect(o_{in}, Step_{OG})$ function collects objects starting from a single $o_{in}$:\\
(1)\hspace*{0.2cm}$collect(o_{in}, Path_{OG}):
	\{ \; O_{i} \subseteq O \; | \; \forall i \in 1...|Path_{OG}|, O_0=\{ o_{in} \}, O_{i} =$\\ 
\hspace*{1cm} $\{ \; \forall o \in O_{i-1}: collect(o, Path_{OG}^i) \; \} \; \}$ {\footnotesize ($Path_{OG}^i$ is the path's i\textsuperscript{th} element)}\\
(2) \hspace*{0.2cm}$collect(o_{in}, Rel_{OG}) =
	\{ \; o \in O \; | \; \\
\hspace*{1cm} ( \; dir(Rel_{OG})=\rightarrow \; \land \; (o_{in}, qual(Rel_{OG}), o) \in O2O \; ) \lor\\ 
\hspace*{1cm}( \; dir(Rel_{OG})=\leftarrow \; \land \; (o, qual(Rel_{OG}), o_{in}) \in O2O \; ) \; \}$\\
(3) \hspace*{0.2cm}$collect(o_{in}, Alt_{OG}):
	\{ \; O_{i} \subseteq O \; | \; \forall i \in 1...|Alt_{OG}|: collect(o_{in}, Alt_{OG}^i) = O_{i} \; \}$
 
\vspace{0.3cm}
For $Path_{OG}$ (1), the function calls itself recursively per step $Path_{OG}^i$, passing objects from the prior step $O_{i-1}$; it yields all returned objects $O_{i}$.
For $Rel_{OG}$ (2), the function returns all objects $o$ with an O2O link to $o_{in}$, based on the step's direction (\textit{dir}) and qualifier (\textit{qual}).
For $Alt_{OG}$ (3), the function also recursively calls itself for each of its steps $Alt_{OG}^i$, passing the same $o_{in}$ each time.

A $\textrm{\textit{Perspective}}(ot_{st}, Path_{OG})$ consists of ``perspective instances'' for each start object, i.e., $\textrm{\textit{Perspective}}(ot_{st}, Path_{OG}) = \{ \; PI(o_{in}, Path_{OG}) \mid objtype(o_{in}) = o_{st} \}$, where $PI(o_{in}, Path_{OG}) = \{ O_{out} \subseteq O \mid O_{out} = collect(o_{in}, Path_{OG}) \; \}$. 


For a given $PI(o_{in}, Path_{OG})$, we can finally define a \textit{trace} as a sequence\footnote{Event ordering will be based on their timestamps (not shown for brevity).} of events linked to the objects in the \textit{PI}, i.e., $\sigma = <e_1, e_2, ..., e_n>$, where $\{ \; e_i \in E \; | \; \exists o \in PI(o_{in}, Path_{OG}): (e_{i}, q, o) \in E2O \}$. 
These traces are serialized in RDF in the same way as shown in Listing~\ref{lst:log_rdf}.

For the simulated P2P log~\cite{p2p_case}, the $Path_{OG}$ will look as follows (using simplified qualifiers for brevity):

\vspace{0.2cm}
{\footnotesize
$( Step_{OG}( \rightarrow , \textrm{``quotation''} ), Step_{OG}( \rightarrow , \textrm{``purchase order''} ),Alt_{OG}($

\noindent \hspace{0.1cm} $Step_{OG}( \rightarrow , \textrm{``invoice receipt''} ), 
Step_{OG}( \rightarrow , \textrm{``payment''} ),
Step_{OG}( \leftarrow , \textrm{``goods receipt''} ) ) )$}

\vspace{0.2cm}
We implemented these definitions using a set of N3 rules to infer the traces~\cite{elkg_repo}. 
We refer to such an extended $\textrm{ELKG}_{ocel2}$ as $ \textrm{ELKG}_{ocel2}^{*ot_{start}}$, e.g., $\textrm{ELKG}_{ocel2}^{*\textrm{\tiny{PR}}}$.

\section{Querying Process Traces}
\label{sec:query_process_traces}

\subsection{Types of trace constraints}
\label{sec:cnstr_types}

Taking inspiration from other trace querying languages~\cite{Schuster-process_querying}, and languages for querying process model repositories~\cite{Polyvyanyy_2019}, we identified an initial set of trace constraints. We provide a summary of their semantics below 
in terms of matching traces.
We loosely group the constraints into two categories, namely \textit{activity occurrence}, i.e., on the (non)-occurrence of activities in traces; and \textit{sequential relations}, i.e., on the temporal ordering between traces' events. For the \textit{activityOccurs} constraints, we also show its set-based (a-b) and cardinality-based (c) versions.
Note that we use $e_a$ to represent an event $e$ with activity $a$.

\vspace{0.15cm}
\noindent \textbf{Activity Occurrence}. 
\begin{enumerate}[topsep=0pt]
    \item \textit{activityOccurs} \textbf{a} -- traces where activity \textit{a} occurs:
    $\{\, \sigma \in L \mid e_a \in \sigma \}$
        \begin{enumerate}
            \item \textit{allActivitiesOccur} \textbf{A} -- traces where \textit{all} activities from set \textit{A} are included:
                $\{\, \sigma \in L \mid \forall a_i \in A: e_{a_i} \in \sigma \}$
            \item \textit{anyActivityOccurs} \textbf{A} -- traces where \textit{any} activity from set \textit{A} is included: \\
                $\{\, \sigma \in L \mid \exists a_i \in A: e_{a_i} \in \sigma \}$
            \item \textit{activityOccursAtLeastNTimes} \textbf{a} $\geq k$ -- traces where activity \textit{a} occurs at least N times:
            $\{\, \sigma \in L \mid | e_{a} \in \sigma | \geq k \}$
        \end{enumerate}
    \item \textit{activityDoesNotOccur} \textbf{a} -- traces where \textit{a} does not occur:
    $\{\, \sigma \in L \mid \neg \exists e_a \in \sigma \}$
    \item \textit{activitiesCoOccurOrNoneOccurs} \textbf{A} -- traces where activities from \textit{A} either all occur or none occur:
    $\{\, \sigma \in L \mid (\forall a_i \in A: \exists e_{a_i} \in \sigma) \vee (\forall a_i \in A: \neg \exists e_{a_i} \in \sigma) \}$
    \item \textit{activitiesDoNotCoOccur} \textbf{A} -- traces where activities from \textit{A} do not occur together; it is possible some of them co-occur (a.k.a. ``conflict''):\\
    $\{\, \sigma \in L \mid \neg ( \forall a_i \in A: \exists e_{a_i} \in \sigma ) \}$
\end{enumerate}
\vspace{0.25cm}
\textbf{Sequential Relations}. 
\begin{enumerate}[topsep=0pt]
    \item \textit{activityOccursAsStart} \textbf{a} -- traces where activity \textit{a} occurs as start.\\
    $\{\, \sigma \in L \mid \exists e_a \in \sigma \wedge \neg \exists e_0 \in \sigma: e_0 \prec e_a \}$
    \item \textit{activityOccursAsEnd} \textbf{a} -- analogous to \textit{activityOccursAsStart}.
    \item \textit{activitiesDirectlyFollow} \textbf{a} \textbf{b} -- traces where activity \textit{a} is directly followed by activity \textit{b}:
    $\{\, \sigma \in L \mid \exists e_{a} \in \sigma \wedge \exists e_{b} \in \sigma: e_a \rightarrow e_b \}$
    \item \textit{activitiesEventuallyFollow} \textbf{a} \textbf{b} -- traces where activity \textit{a} is eventually followed by activity \textit{b}:
    $\{\, \sigma \in L \mid \exists e_{a} \in \sigma \wedge \exists e_{b} \in \sigma: e_a \prec e_b \}$
    \item \textit{activitiesAlwaysPrecede} \textbf{A} \textbf{B} -- traces where all activities from \textit{A} always precede the activities from \textbf{B} (a.k.a. ``total-causal''):\\
    $\{\, \sigma \in L \mid \forall a_i \in A \wedge \forall b_i \in B: e_{a_i} \in \sigma \wedge e_{b_i} \in \sigma \wedge e_{a_i} \prec e_{b_i} \}$
    
\end{enumerate}

\subsection{Implementing trace constraints in N3}
\label{sec:seq_rel_n3}
We implemented the constraints from Section~\ref{sec:cnstr_types} in the N3 rule language. 
Trace queries (Section~\ref{sec:queries}) can refer to these rules to find particular traces.
We note that these implementation patterns, which may apply to other rule languages as well, can be used to implement other constraints not yet considered here.

\vspace{-0.5cm}
\subsubsection{Activity Occurrence.} 
Below, we show the N3 rule for \textit{activityOccurs}:

\lstset{caption={\textit{activityOccurs} in N3}, label=lst:activityOccurs}
\begin{lstlisting} 
{ ?t pq:activityOccurs ( ?a ?e ) } 
  <= { ?e tr:in ?t . ?e tr:activity ?a } .
\end{lstlisting}

An activity \textit{?a} occurs in the form of event \textit{?e} within trace \textit{?t}, if the trace includes (\textit{tr:in}) the event, and the event is about (\textit{tr:activity}) the activity. This rule refers to the RDF log structure as illustrated in listing~\ref{lst:log_rdf}.

To implement set-based or cardinality-based versions of a constraint,
we refer to the original constraint's rule combined with N3 builtins.
We rely on \textit{list:in} to implement the existential (\textit{any}) set-based version. For \textit{activityOccurs}:
\lstset{caption={\textit{anyActivityOccurs} in N3}, label=lst:anyActivityOccurs}
\begin{lstlisting} 
{ ?t pq:anyActivityOccurs ?activities }
  <= { ?a list:in ?activities . ?t pq:activityOccurs ( ?a ?e ) } .
\end{lstlisting}

The rule checks whether one of the activities \textit{?a} from the \textit{?activities} list (\textit{list:in} builtin) occurs in trace \textit{?t}, as checked by the prior \textit{pg:activityOccurs} rule.

To implement the universal (\textit{all}) set-based version of a constraint, we rely on builtins that implement a Scoped Negation As Failure (SNAF)\footnote{N3
allows for the assumption that a particular scope contains all relevant statements.}:

\lstset{caption={\textit{allActivitiesOccur} in N3}, label=lst:allActivitiesOccur}
\begin{lstlisting} 
{ ?t pq:allActivitiesOccur ?activities }
  <= { ?t a tr:Trace .
       ( { ?a list:in ?activities } 
         { ?t pq:activityOccurs ( ?a ?e ) } ) 
       log:forAllIn _:t } .
\end{lstlisting}

The \textit{log:forAllIn} SNAF builtin~\cite{van_woensel_notation3_2023} accepts a subject list (lines 3-4) with two clauses (graph terms):
for every match of the first clause, the second clause must also hold.
Here, the builtin checks that for every activity \textit{?a} in the \textit{?activities} list (\textit{list:in}), the activity also occurs in trace \textit{?t} (\textit{pg:activityOccurs} rule).

To implement a cardinality-based version, we rely on \textit{log:collectAllIn}:

\lstset{caption={\textit{activityOccursAtLeastNTimes} in N3}, label=lst:activityOccursAtLeastNTimes}
\begin{lstlisting} 
{ ?t pq:activityOccursAtLeastNTimes ( ?a ?atLeastN ) }
<= { ?t a tr:Trace .
     ( ?e { ?t pq:activityOccurs ( ?a ?e ) } ?evts ) 
         log:collectAllIn _:t .
     ?evts list:length ?n .
     ?n math:notLessThan ?atLeastN } .
\end{lstlisting}

The \textit{log:collectAllIn} builtin collects all values that match a given clause into a list, based on a subject list. Here (line 3), it will collect all events \textit{?e} of which the activity \textit{?a} occurs within the given trace \textit{?t} (\textit{pg:activityOccurs} rule), and add them to the list \textit{?evts}. Next, the rule checks whether the length of list \textit{?evts}, i.e., the number of occurrences of \textit{?a} in trace \textit{?t}, is greater than or equal to (\textit{notLessThan}) number \textit{?atLeastN}. Other cardinality restrictions (e.g., \textit{AtMost}) can be represented with the appropriate math builtin.

To check whether an activity \textit{does not} occur (\textit{activityDoesNotOccur}):

\lstset{caption={\textit{activityDoesNotOccur} in N3}, label=lst:activityDoesNotOccur}
\begin{lstlisting} 
{ ?t pq:activityDoesNotOccur ?a }
  <= { ?t a tr:Trace .
       _:t log:notIncludes { ?e tr:activity ?a . ?e tr:in ?t } .
\end{lstlisting}

If there does not exist (\textit{log:notIncludes} SNAF builtin) an event about the activity within \textit{?t}, then an activity \textit{?a} does not occur within a trace \textit{?t}.

\vspace{-0.3cm}
\subsubsection{Sequential Relations.} Similar rules and builtins can be used to implement several of the sequential relation constraints. 

We rely on rule recursion to implement \textit{activitiesEventuallyFollow}:

\lstset{caption={\textit{activitiesEventuallyFollow} in N3}, label=lst:activitiesEventuallyFollow}
\begin{lstlisting} 
{ ?t pq:activitiesEventuallyFollow ( ?a ?b ) }
  <= {  ?t pq:activityOccurs ( ?a ?e ) .
        ?t pq:eventPrecedesActivity ( ?e ?b ) } .

# navigate sequential relations recursively
{ ?t pq:eventPrecedesActivity ( ?e ?b ) }
  <= {  ?e tr:activity ?e_a . ?e_a log:notEqualTo ?b . # not about b
        ?e tr:next ?e_n  # next event e_n
        ?t pq:eventPrecedesActivity ( ?e_n ?b ) } . # search from e_n

{ ?t pq:eventPrecedesActivity ( ?e ?b ) } 
  <= {  ?e tr:activity ?b } . # event is about b
\end{lstlisting}

The first rule (lines 1-3) gets the event $?e$ for activity $?a$, using the \textit{activityOccurs} constraint. The second rule (lines 5-9) checks whether this event $?e$ precedes activity $?b$. It recursively navigates sequential relations, starting from the given event, for as long as the next event's activity is not $?b$. If so, rule 3 (lines 11-12) ensures the rule finally resolves.
Finally, our online repo~\cite{elkg_repo} includes an example query that checks the duration between events using N3 builtins.

Our approach is extensible, since extra constraints can easily be implemented in N3; e.g., retrieving events occurring between two activities~\cite{bpmn-q_2007} can be similarly implemented using recursive rules.
Furthermore, multiple constraint versions can be combined (e.g., \textit{allActivitiesOccurAtLeastNTimes}), 
or set-based or cardinality-based versions of other constraints can be similarly implemented.
We did this for a number of constraints; refer to our online repo~\cite{elkg_repo} for all N3 rules.

\subsection{Querying Entities using Rules and Other Constraints}
\label{sec:queries}
Queries for finding traces, activities, or events from the event log
can utilize the backward-chaining rules from the prior section. 
E.g., this forward-chaining rule ($\Rightarrow$), acting as a query, returns all traces where purchase requisitions were created but not approved (i.e., with activity ``Create Purchase Requisition'' but without ``Approve Purchase Requisition''; a.k.a. ``maverick buying''):

\lstset{caption={Example ``maverick buying'' query for P2P use case.}, label=lst:mav_buying_query}
\begin{lstlisting} 
{   ?t pq:activityOccurs ( "Create Purchase Requisition" ?e ) .
  ?t pq:activityDoesNotOccur "Approve Purchase Requisition" .
} => { pq:result pq:entry ?t } .
\end{lstlisting}

N3, as a general-purpose rule language, offers the expressivity to instantiate constraints in multiple ways.
Here, by using the same variable \verb|?t| as subject, the query will only return traces adhering to both constraints.
For instance, these identified traces, with events covering PR creation to final payment, can be investigated to find the reasons and/or effects (e.g., duration, final cost) of maverick buying.
E.g., another example query (see online repo~\cite{elkg_repo}) leverages the total duration of extracted traces in {\footnotesize $ELKG_{OG}^{*\textrm{\tiny{PR}}}$}, i.e., from PR to payment.

A more comprehensive example for the sepsis use case, which leverages sequential relations, will return all traces with at least one lactic acid and CRP lab test event; a directly-follows relation between admission to NC and IC units; and concerning a patient with a suspected infection and aged 65 or over:

\lstset{caption={Example query with multiple constraints for sepsis use case.}, label=lst:large_sepsis_query}
\begin{lstlisting} 
{   ?t pq:activityOccurs ( :LacticAcid ?la ) .
  ?t pq:activityOccurs ( :CRP ?crp ) .
  ?t pq:activitiesDirectlyFollow ( :Admission_NC :Admission_IC ?r ).
  ?t pq:activityOccurs ( :ER_Registration ?reg ) .
  ?reg :Age ?age . ?age math:notLessThan 65 .
  ?reg :InfectionSuspected true } 
=> { pq:result pq:trace ?t} .
\end{lstlisting}


\section{Preliminary Performance Evaluation}
\label{sec:eval}

We conducted a preliminary performance evaluation on the well-known CCEL sepsis event log~\cite{mannhardt_sepsis} with 1050 process traces
; and the OCEL2 P2P simulated dataset with 14671 events, 10 event types, 9543 objects and 7 object types ~\cite{p2p_case}.
Experiments were performed on a Macbook Pro with an Apple M1 Pro chip and 32Gb memory. Performance results were obtained over an average of 5 runs.
We used Python v.3.11.5 with pandas v.2.0.3, rdflib v.7.0.0, pm4py v.2.7.11.12, and the \textit{eye}~\cite{eye} N3 reasoner v10.17.3 for executing N3 rules.

\vspace{0.2cm}
\noindent \textbf{Converting Event Logs into ELKG}. Converting the sepsis CCEL into RDF took around 3,9s on average.
Converting the P2P OCEL2 into RDF took around 10,5s on average. Extracting perspective-based traces led to an {\footnotesize $ELKG_{OCEL2}^{*\textrm{\tiny{PR}}}$} with 927 traces; this step is currently not optimized and took around 2,7min. 

\vspace{0.2cm}
\noindent \textbf{Querying Process Traces}.
We executed a set of queries, each of which encapsulating a single trace constraint, on the sepsis event log. Loading the ELKG into the N3 reasoner took around 410ms on average; executing each query took around 42ms on average.
The sepsis query with multiple constraints, shown in Listing~\ref{lst:large_sepsis_query}, took around 120ms on average.

For the P2P event log, we formulated a number of queries to identify ``special behaviors'' as mentioned by the dataset authors~\cite{p2p_case}, i.e., maverick buying (Listing~\ref{lst:mav_buying_query}), duplicate payments, and lengthy (approval) processes. 
Here, loading the {\footnotesize $ELKG_{OCEL2}^{+PR}$} into the N3 reasoner took around 1240ms on average; executing each query took around 90ms on average.
 
We refer to our online repo~\cite{elkg_repo} for all queries and results. We also provide a Jupyter notebook for manually testing the conversion steps and queries.

\section{Related Work}
\label{sec:rel_work}
Most approaches for querying process traces~\cite{Schuster-process_querying,Beheshti-process_querying,Vogelgesang2022} and process model repositories~\cite{Polyvyanyy_2019,bpmn-q_2007}
propose a separate domain-specific language~\cite{Schuster-process_querying,Vogelgesang2022,Polyvyanyy_2019,bpmn-q_2007}; while this makes queries more concise and likely more performant, it comes at the cost of expressivity and extensibility.
Beheshti et al.~\cite{Beheshti-process_querying} extended an existing query language (SPARQL), which similarly allows restricting arbitrary entity properties and relations.
However, in contrast to N3, SPARQL itself lacks the expressivity to represent trace constraints; a separate regular expression processor searches for event paths. 
Kobeissi et al.~\cite{Kobeissi-process_querying} 
issues queries over property graphs of event logs, retrieving nodes based on relations between them; however, there is no particular support for querying activity occurrences or sequential relations.

On the extraction of traces from an OCEL2,  
Berti et al.~\cite{berti_oc-pm_2023}
mention the possibility to filter events given a set of objects, but do not elaborate on how to identify these objects. 
Fahland~\cite{Fahland2022} 
leverages an Eevent Knowledge Graph (EKG) to infer object type relations: if a single event involves both objects \textit{I2} (invoice) and \textit{O2} (order),
a relation can be inferred between the object types, i.e., {\footnotesize $R(invoice, order)$}. 
If a following event further involves \textit{I2} (invoice) and \textit{P2} (payment), a transitive relation {\footnotesize $R(order, payment)$} can be inferred. 
Clearly, it requires domain knowledge to decide which relations to infer~\cite{Fahland2022}.
In case an object graph (based on O2O relations in an OCEL2 log) is not available, this may be a viable approach to infer implicit object relations. 

\section{Conclusions and Future Work}
\label{sec:concl_fut_work}

This paper showed how different types of event logs, i.e., CCEL and OCEL2, can be converted into a semantic trace-based ELKG. 
To obtain meaningful traces from an OCEL2, we can ``flatten'' the log based on an arbitrary perspective, defined in terms of an object path through the ELKG.
Within an ELKG, we showed how occurrence and sequential relation constraints can be implemented in N3 in an expressive and extensible way.
Regarding performance, converting the logs into ELKG takes less than 10-11s, and is a one-time effort; however, extracting traces from {\footnotesize $ELKG_{OCEL2}$} is relatively slow, around 3 min for the P2P log.
For querying traces, while loading an ELKG into \textit{eye} is a non-trivial task (around 400-1300ms), executing queries is more performative (under 125ms).

There remain many avenues for future work. 
We aim to optimize the extraction of traces and loading/querying of ELKG.
We will try out other OCEL2 to assess the utility of ``perspectives'' and how they are currently constructed. 
We are targeting more constraints (e.g., retrieving sequences of events~\cite{bpmn-q_2007}), and support aggregate statistics (e.g., percentage of retrieved traces). 
We aim to make ELKG from CCEL and OCEL2 more homogenous: e.g., in the sepsis dataset, attributes are associated with events; in an OCEL2 log, these attributes would likely be represented using objects.
Ideally, an ELKG would represent this data in the same way.
Sequential relations are currently represented using predicates; a more flexible and thus desirable alternative is reifying this relation, but currently this drastically reduces performance.
Finally, we aim to support partial temporal ordering as opposed to the total ordering assumed in this paper.



%
%
\bibliographystyle{splncs04}
\bibliography{paper}

\end{document}